\documentclass[]{aa}
\usepackage[T1]{fontenc}
\usepackage[utf8]{inputenx}
\usepackage{lmodern}
\usepackage{graphicx}
\usepackage{natbib}
\bibpunct{(}{)}{;}{a}{}{,} 
\setcitestyle{round}
\usepackage{ulem}
\usepackage{textcomp}
\usepackage{gensymb}
\usepackage{longtable}
\usepackage{threeparttable}
\usepackage{multicol}
\usepackage{multirow}
\usepackage{lipsum}
\usepackage{float}
\usepackage{todonotes}
\usepackage[Symbol]{upgreek}
\usepackage{amsmath}
\usepackage{etoolbox}
\usepackage{txfonts}
\usepackage{url}
\usepackage{xcolor}
\usepackage{mathtools}
\usepackage[most]{tcolorbox}
\usepackage[colorlinks=true,allcolors=blue]{hyperref}
\usepackage{booktabs, threeparttable}

\usepackage[symbol]{footmisc}

\usepackage{array}
\newcolumntype{L}[1]{>{\raggedright\let\newline\\\arraybackslash\hspace{0pt}}m{#1}}
\newcolumntype{C}[1]{>{\centering\let\newline\\\arraybackslash\hspace{0pt}}m{#1}}
\newcolumntype{R}[1]{>{\raggedleft\let\newline\\\arraybackslash\hspace{0pt}}m{#1}}

\begin{document}
 
\title{Revisiting the alignment of radio galaxies in the ELAIS-N1 field\textcolor{blue}{$^\star$}}

\author{M. Simonte \inst{1}, H. Andernach \inst{2}, M. Br\"uggen \inst{1}, P. N. Best\inst{3}, E. Osinga\inst{4}}

\institute{Hamburger Sternwarte, University of Hamburg, Gojenbergsweg 112, 21029 Hamburg, Germany; marco.simonte@hs.uni-hamburg.de
\and Th\"uringer Landessternwarte, Sternwarte 5, D-07778 Tautenburg, Germany; permanent address: Depto. de Astronomia, Univ. de Guanajuato,
Callej\'on de Jalisco s/n, Guanajuato, C.P. 36023, GTO, Mexico; heinz@ugto.mx.
\and Institute for Astronomy, University of Edinburgh, Royal Observatory, Blackford Hill, Edinburgh, EH9 3HJ, UK
\and Leiden Observatory, Leiden University, PO Box 9513, 2300 RA Leiden, The Netherlands\\}

 \authorrunning{M. Simonte et al.}
 \titlerunning{Alignment of RGs in the ELAIS-N1 field}

\date{Accepted ???. Received ???; in original form ???}

\abstract
  {}
   {Previous studies reported an alignment of the major axes of radio galaxies on various angular scales. Here, we study the alignment of radio galaxies in the ELAIS-N1 Low Frequency ARray (LOFAR) deep field, which covers an area of 25 $\rm deg^2$.}
   {The low noise level of about 20$ \rm ~ \micro Jy/beam$ of the LOFAR deep field observations at 150 MHz enabled the identification of 447 extended ($> 30 \rm ''$) radio galaxies for which we have measured the major axis position angle. We found that 95\% of these sources have either photometric or spectroscopic redshifts, which we then used for a three-dimensional analysis.}
   {We show the distribution of the position angles of radio galaxies in the ELAIS-N1 field and perform multiple statistical tests to check whether the radio galaxies are randomly oriented. We found that the distribution of position angles is consistent with being uniform. Two peaks around position angles of 50 and 140$\rm~ deg$ are spurious and are not caused by an alignment, as shown by a 3D analysis. In conclusion, our results do not support a 2D or 3D alignment of radio galaxies on scales smaller than  $\sim 4 \rm ~ deg$.}
   {}



\maketitle

\section{Introduction}

\footnote[0]{\textcolor{blue}{$^\star$} The full Table~\ref{tab:ERGs_list} is only available in electronic form
at the CDS via anonymous ftp to \href{cdsarc.cds.unistra.fr (130.79.128.5)}{cdsarc.cds.unistra.fr (130.79.128.5)}
or via \href{https://cdsarc.cds.unistra.fr/cgi-bin/qcat?J/A+A/}{https://cdsarc.cds.unistra.fr/cgi-bin/qcat?J/A+A/}
}

The cosmological principle is an assumption in modern cosmology which states that the Universe is (statistically) isotropic and homogeneous on suitably large scales ($\gtrsim 100 \rm ~ Mpc)$. Multiple observations have investigated the degree of anisotropy in the cosmic microwave background \citep{Bennett1996, Hansen2004, Planck2016, Planck2020} confirming the principle of homogeneity and isotropy of the Universe. However, several authors have reported an intriguing alignment of the linear polarisation of quasars \citep{Stockman1979,Hutsemekers1998,Hutsemekers2001, Jain2004,Cabanac2005,Pelgrims2014,Slagter2021, Friday2022}. Interestingly, they found an alignment mainly occurring in groups of 10-30 objects and on potentially Gpc scales.\newline
Some other studies focused on the alignment of radio galaxy jets \citep[e.g.,][]{Sanders1984,Kapahi1985, West1991,Joshi2007, Tiwari2013}, some of which support a possible departure from the cosmological principle. \citet{Taylor2016} studied the spatial distributions of the major axis position angle of radio galaxies in the ELAIS-N1 Giant Metrewave Radio Telescope (GMRT, \citealt{GMRT95}) deep field. They claimed the existence of a 2D alignment around $PA\sim140^{\circ}$ over an area of $\sim1.7 \rm~deg^2$. However, for lack of the redshifts of the host galaxies, they did not perform a 3D analysis. The first attempts to detect an alignment on larger scales were made by \citet{Contigiani2017} and \citet{Panwar2020} who used catalogue data from the Faint Images of the Radio Sky at Twenty-cm \citep[FIRST,][]{FIRST1995, Helfand2015} and the TIFR GMRT Sky Survey \citep[TGSS,][]{Intema2017}. They detected a signal over a scale smaller than 2$^{\circ}$, but did not find strong evidence for a 3D alignment. For the first time, \citet{Blinov2020} explore the alignment of parsec-scale jets finding that their radio sources do not show any global alignment. However, \citet{Mandarakas2021}, with a similar but larger sample, detected a strong signal of an alignment of parsec-scale jets in multiple regions of the sky. Nevertheless, the redshift distribution of their sources spans a wide range, $0 < z \lesssim 1.5$ Most recently, \citet{Osinga2020} searched for alignment using 7555 extended sources from the first data release of the Low Frequency ARray Two metre Sky Survey \citep[LoTSS,][]{DR12019}. However, despite their use of host redshifts, they could only detect a 2D alignment of the position angles of the radio galaxies over a scale of 5$^\circ$ and could not exclude the possibility that the signal arises from systematic effects. \newline
Although multiple studies have now presented evidence for a 2D or 3D alignment, an explanation for such a phenomenon is lacking. \citet{West1991}, \citet{Hutsemekers2014} \citet{Pelgrims2016} found an alignment between the radio and optical emissions from active galactic nuclei (AGN)  and the surrounding large-scale structure. Moreover, \citet{Malarecki2013, Malarecki2015} showed that giant radio galaxies \citep{Willis1974} have a tendency to grow in a direction perpendicular to the major axes of galaxy overdensities. However, the connection between the orientation of radio galaxy jets and the large-scale structure is unclear. \newline
In this paper, we revisit the alignment of radio galaxies jets in the ELAIS-N1 field. We make use of photometric redshifts of the host galaxies to perform a 3D analysis. \newline
The outline of this paper is as follows: In Sec.~\ref{sec:methods} we explain how we built our catalogue of extended radio galaxies (ERGs) and how we measured the orientation of the radio galaxies. In Sec.~\ref{sec:results} we present our results of the 2D and 3D analysis. In Sec.~\ref{sec:discussion}, we discuss our results in the context of theoretical and observational work on the orientation of radio galaxies and give our summary.\newline
Throughout this work we adopt a flat $\rm \Lambda CDM$ cosmology with \textit{$H_0$} = 70 $\rm ~ km ~ s^{-1} ~ Mpc^{-1}, ~ \Omega_m = 0.3, ~ \Omega_{\Lambda} = 0.7$.

\section{Methods}
\label{sec:methods}

We inspected the ELAIS-N1 LOw-Frequency ARray \citep[LOFAR,][]{Lofar2013} deep field \citep{Sabater2021}. With an effective observing time of 163.7 h, it reaches a root mean square noise level at 150 MHz lower than $30 \rm ~ \mu Jy ~ beam^{-1}$ across the inner 10 $\rm deg^2$ and below $20 \rm ~ \mu Jy ~ beam^{-1}$ in the very centre. The ELAIS-N1 LOFAR Deep Field (ELDF) is centred on $ \rm 16h11m00s +55^{\circ}00\rm '00\rm '' (J2000)$ and it covers an area of about 25 $\rm deg^2$. The 6$\rm "$ resolution of the radio image ensures a robust classification of the sources and, most importantly, the identification of the hosts and radio features such as jets and hotspots. 

\subsection{The sample of extended radio galaxies}

We searched for all the ERGs with a largest angular size (LAS) larger than $\sim 30\rm"$ within an area of $\sim$ 25 $\rm deg^2$. We measured the LAS as the distance between the opposite ends of the ERGs. However, this method can overestimate the size of the Fanaroff-Riley type II \citep[FRII,][]{FR1974} as commented in \citet{Kuzmicz2021}. Thus, for such ERGs, we measured the LAS as the distance between the two hotspots, whenever identified on the VLA Sky Survey images \citep{Lacy2020}. The radio position angles (RPAs) were manually measured (by using Aladin\footnote{ \href{https://aladin.cds.unistra.fr}{https://aladin.cds.unistra.fr}}, \citealt{Bonnarel2000}) in the range [0,180) degrees as the angle between the source's major axis and the local meridian, from N through E. For straight (or only slightly bent) FRI and FRII, the RPA is either that of the inner jets (FR\,I) or that of the direction connecting the two hotspots (FR\,II). In the case of bent sources (e.g., Wide-Angle-Tailed RGs), measuring the RPA is less trivial. For such cases, we measured the RPA in the vicinity of the core where usually the jets are not bent yet and flagged them as uncertain measurements. We carefully avoided measuring the RPA of overlapping sources unless the morphology of the ERGs was very clear. \newline
A large number of optical and infrared surveys, such as the Wide-Field Infrared Survey Explorer \citep[WISE,][]{Cutri2012,Cutri2013,unWISE2019,Marocco2021}, the Sloan Digital Sky Survey \citep[SDSS,][]{York2000}, the Legacy survey \citep[][]{Legacy2019} and the Panoramic Survey Telescope and Rapid Response System \citep[Pan-STARRS,][]{panstarss} enabled the identification of the host galaxies \citep[see][for further details on the host identification and radio source classification]{Kondapally2021,Andernach2021,Simonte2022}. We looked for available redshifts (either spectroscopic or photometric) in multiple catalogues such as \citet{RR2013}, \citet{Bilicki2014}, \citet{Bilicki2016}, \citet{Beck2016},\citet{Duncan2021}, \citet{Beck2021}, \citet{Zhou2021}, \citet{Wen2021} and \citet{Duncan2022}. If for a single source multiple photometric redshifts were available, we computed their mean and error by taking the standard deviation of the various redshifts. For spectroscopic redshifts, we do not report errors since they are generally more accurate (typical errors are usually around 0.00015) than the precision we can achieve on the linear size given our errors in measuring the angular size. For 15 optically very faint host or infrared-only detected host
galaxies, no redshift estimate was available. The deepest full-sky catalogue is the \citet{Zhou2021} DESI DR9 photometric redshift catalogue (a deeper catalogue from \citet{Duncan2021} exists over the inner 7$\rm deg^2$ of the ELAIS-N1 field). In \citet{Zhou2021} the faintest galaxies have a maximum redshift of around 1.3. Thus, we assumed a redshift in the range of 1.1-1.5 for those host galaxies without redshift listed in the literature. This assumption will not affect our analysis as we will use only those sources with a redshift reported in the literature for the 3D analysis. \newline
We found 447 ERGs for which we provide redshift, LAS, largest linear size (LLS) and RPA. We show some of our ERGs in Table~\ref{tab:ERGs_list} and the full list will be made available at the CDS and through the VizieR service\footnote{ \href{https://vizier.cds.unistra.fr}{https://vizier.cds.unistra.fr}} \citep{Ochsenbein2000}. 
To test the alignment in the region inspected by \citet{Taylor2016}, we located all sources these authors had used (their Fig.~2) and measured their RPAs. Some of these RGs have an angular size smaller than $30 \rm ''$. The resolution of 6" of the LOFAR images does not enable reliable measurement of the RPA of the smallest sources and we flagged these measurements as uncertain. We had to discard 9 RGs used by \citet{Taylor2016} as 8 of them are separate sources and one is a spiral galaxy (see Appendix~\ref{appendix}). However, we were able to add 24 more ERGs within the sky area they studied that we were able to identify using the LOFAR data.
In Tab.~\ref{tab:comparison} we compare our sample with previous lists of RGs used for the RPA analysis. In this work, we analysed a field $\sim$ 10 times larger than that of \citet{Taylor2016}, but much smaller
than those used by \citet{Contigiani2017}, \citet{Panwar2020} and \citet{Osinga2020}. Nevertheless, our sample has the largest RGs sky density in the central region ($241.5^\circ < RA < 243.75^\circ, 53.9^\circ < DEC < 55.2^\circ$), which is reported in the last row, while the second-last row shows the RGs sky density considering the full ELDF. 

 \begin{center}
 \begin{table}
    \centering
    \begin{tabular}{p{2.7cm}p{0.5cm}p{0.8cm}p{0.8cm}p{1.8cm}}
    \hline
      (1) & (2) & (3) & (4) & (5) \\
      Survey & Freq. & RMS & N of & RGs density \\ 
      & GHz & mJy/b & RGs & deg$^{-2}$   \\ \hline
      Taylor$^1$ & 0.61 & 0.01 & 65 & 38.2 \\
      FIRST$^2$ & 1.4 & 0.15 & 30059 & 4.3 \\ 
      FIRST$^3$ & 1.4 & 0.15 & 18775 & 1.9 \\
      LoTSS$^4$ & 0.15 & 0.07 & 7555 & 17.8 \\
      ELDF$^5$ & 0.15 & 0.03 & 447 & 17.9 \\
      ELDF-C$^6$ & 0.15 & 0.02 & 78 & 45.9 \\
 \hline
    \end{tabular}
    \caption{Comparison between our catalogue and previous samples. References: 1-\citet{Taylor2016}, 2-\citet{Contigiani2017}, 3-\citet{Panwar2020}, 4-\citet{Osinga2020}, 5,6-this work: ELDF-C refers to the central region of the ELDF ($241.5^\circ < RA < 243.75^\circ, 53.9^\circ < DEC < 55.2^\circ$).}
    \label{tab:comparison}
\end{table}
\end{center}

\subsection{Statistical tests}
\label{sec:tests}

We performed multiple tests to assess the (non-)uniformity of the RPA distribution. Different methods have been used in past analyses to study the distribution of the orientation of RGs. 
We use five different tests for (non-)uniformity of angles: \newline
1. The Kolmogorov-Smirnov (KS) test compares the underlying distribution of the sample of the RPA against a given distribution, which in our case is a uniform distribution. The null hypothesis is that the two distributions are identical and the closer the p-value is to zero the more confident we are in rejecting the null hypothesis. A common threshold used to reject the null hypothesis of the two distributions being drawn from the same population is a p-value p$<$0.05, which means that there is only a 5\% chance that the two samples are in fact drawn from the same population.\newline
2. Pearson's $\chi^2$ test for uniformity tests the null hypothesis stating that the frequency distribution of certain events observed in a sample is consistent with a particular theoretical distribution (in our case a uniform one). As with the KS test, the smaller the p-value the more likely it is that the two distributions are different. This test is performed with binned data and, in our case, we used 18 bins which are 10$^\circ$ wide.\newline
3. Our set of RPAs belongs to the category of circular data \citep{Fisher1993} which are fundamentally different from 
linear data due to their periodic nature. The Rayleigh test \citep{Mardia2000} assesses the uniformity of circular data. To this end, this test compares the test statistic of the unit vector, resulting from the sum of all the vectors pointing towards the different angles of the sample, with the same statistics estimated from a uniformly distributed sample. The null hypothesis of such test is that the data are uniformly distributed over the circle. The test statistic is the mean resultant length of the unit vector and it is  defined as 

\begin{equation}
    \bar{R} = \frac{1}{n}\left[ \left( \sum_{i=1}^n{\cos\theta_i}\right)^2 + \left( \sum_{i=1}^n{\sin\theta_i}\right)^2 \right]^{1/2},
\label{eq:R}
\end{equation}
\noindent
where n is the size of the sample and the angles $\theta_i$ are the RPAs multiplied by two since these are orientations (axial vectors) in the
range [0$^\circ$, 180$^\circ$) while the Rayleigh test is performed on the range [0$^\circ$, 360$^\circ$). $\bar{R}$ can range from 0 to 1. This statistic is zero for a uniform distribution, thus it is reasonable to reject uniformity when $R$ is large. It is worth mentioning that this test is not sensitive to non-uniform distributions that have $\bar{R}$ = 0. An example is a bimodal distribution with two peaks that are 180$^\circ$ apart as every vector pointing towards a certain direction is cancelled by a vector pointing along the opposite direction. This issue can mildly affect our analysis since the major peaks in our distributions of the RPAs are 180$^\circ$ apart once the RPAs are multiplied by two (see Sec.~\ref{sec:results} below).\newline 
4. The semi-variance \citep{Cressie} is a statistical tool used in spatial analysis to measure the dispersion of a certain variable on different scales. It is defined as follows:

\begin{equation}
    \gamma(d) = \frac{1}{2m(d)} \sum_{i=1}^{m(d)} \left[ s(x_i) - s(x_i+d) \right]^2,
\label{eq:semi-variance}
\end{equation} 
\noindent
where $m(d)$ is the number of pairs separated by an (angular) distance in the range [$d$, $d+\delta d]$ (we used $\delta d$ = 0.2$^\circ$) and $s$ is the variable measured at the vector location $x_i$ and in our case is the RPA of the ERGs. The semi-variance is constant over all angular scales when the distribution of the variable $s$ is uniform. A value for the semi-variance smaller than what is predicted by a uniform distribution at a certain scale indicates an alignment of the ERGs. On the other hand, a larger semi-variance suggests a larger dispersion than expected from a random distribution, indicating that no alignment is present on that scale. We performed a simple Monte-Carlo simulation to infer the value of the semi-variance of randomly distributed ERGs on different angular scales. We generated 447 (which is the size of our sample) random angles uniformly distributed in the range $\left[ 0,180 \right)$ which have the same spatial distribution of the ERGs in our sample and we computed the semi-variance on different angular scales. We repeated the operation 10000 times and then averaged the semi-variance values on the different scales. We folded the data in circularity to take into account the periodicity of the RPAs. On every scale, we obtained a constant semi-variance of 0.82, consistent with the result from \citet{Taylor2016}. The error on the semi-variance, $\sigma_{\rm SM}$ was estimated by calculating the standard deviation of the 10000 values on each angular scale. \newline
\noindent
5. Finally, we probed the alignment of the ERGs at different angular scales using the dispersion measure analysis \citep{Jain2004}. The dispersion measure is defined as the inner product between a certain position angle $\theta$ and the RPAs, $\theta_k$, of the $n$ closest sources to a certain $i$-th ERG (including the source itself) and it is an indication of the alignment of the ERGs. Following \citet{Jain2004}, \citet{Contigiani2017} and \citet{Osinga2020}, it can be shown that the maximum dispersion measure around the source $i$ is

\begin{equation}
    D_{i,n}|_{\rm max} = \frac{1}{n} \left[ \left( \sum_{k=1}^{n}{\cos(\theta_k)} \right)^2 + \left(\sum_{k=1}^{n}{\sin(\theta_k)} \right)^2  \right]^{1/2}.
\label{eq:dispersion_measure}
\end{equation}
The closer $D_{i,n}|_{\rm max}$ is to 1, the more aligned the $n$ galaxies are. The statistic, $S_n$ used to test the (non-)uniformity of the distribution of the RPAs is the average of the $D_{i,n}|_{\rm max}$ calculated for each source of the sample. This statistic computed from our dataset is compared to the same statistics coming from Monte-Carlo simulated samples, $S_{\rm n,MC}$. To compute $S_{\rm n,MC}$ we generated 447 randomly oriented ERGs with the same spatial distribution of our sources and followed the formalism described in \citet{Jain2004}, \citet{Contigiani2017} and \citet{Osinga2020}. We repeated the calculation of $S_{\rm n,MC}$ 10000 times and estimate the average, $\langle S_{\rm n,MC}\rangle$, and the error, $\sigma_{\rm n,MC}$, as the standard deviation of 10000 generated statistics. The significance level for rejecting the null hypothesis that a sample of ERGs is randomly oriented is found through a one-tailed significance test, expressed as:

\begin{equation}
    SL = 1 - \Phi \left(\frac{S_{\rm n} - \langle S_{\rm n,MC}\rangle}{\sigma_{\rm n,MC}} \right),
\label{eq:onetail_test}
\end{equation}
where $\Phi$ is the cumulative normal distribution function. The closer the significance level is to 0 the more confident we are in rejecting the hypothesis of uniformity. Since the number of nearest neighbours can be translated to an angular scale extending to the $n$-th nearest neighbour, we can probe multiple angular scales varying $n$. To do so, we calculated the maximum angular distance between the relevant ERG and the n-th closest neighbour and took the median value among the 447 sources. The same analysis can be implemented considering the 3D position of the ERGs to test whether a 3D alignment, i.e. between sources that are
physically close to each other, is present. We approximated the redshift of the source with the average redshift estimated for each ERGs without taking into account the error and we did not include those sources without a redshift value reported in the literature. The uncertainties of some redshift estimations might mildly affect the analysis: in fact, while ERGs with $z < 1$ have a redshift error of about 0.05, for more distant sources, which represent 30\% of our sample, the error increases to 0.2. We then converted the redshift to comoving distance and measured the 3D comoving distance between all the ERGs in our sample. \newline
Moreover, \citet{Jain2004} verified that the variance of the statistic $S_n$ is inversely proportional to the size of the sample which means that, compared to \citet{Contigiani2017} and \citet{Osinga2020} who used much larger samples, we are more affected by the shot noise. 

\section{Results}
\label{sec:results}

In this section, we present the distribution of the RPAs in the ELAIS-N1 field. We initially focus on the inner region studied by \citet{Taylor2016} and then expand the analysis to the entire ELDF.

\subsection{Alignment in the central part of ELAIS-N1}
\label{sec:Taylor_alignment}

\begin{figure}[]
        \centering
        \includegraphics[width= 0.50\textwidth]{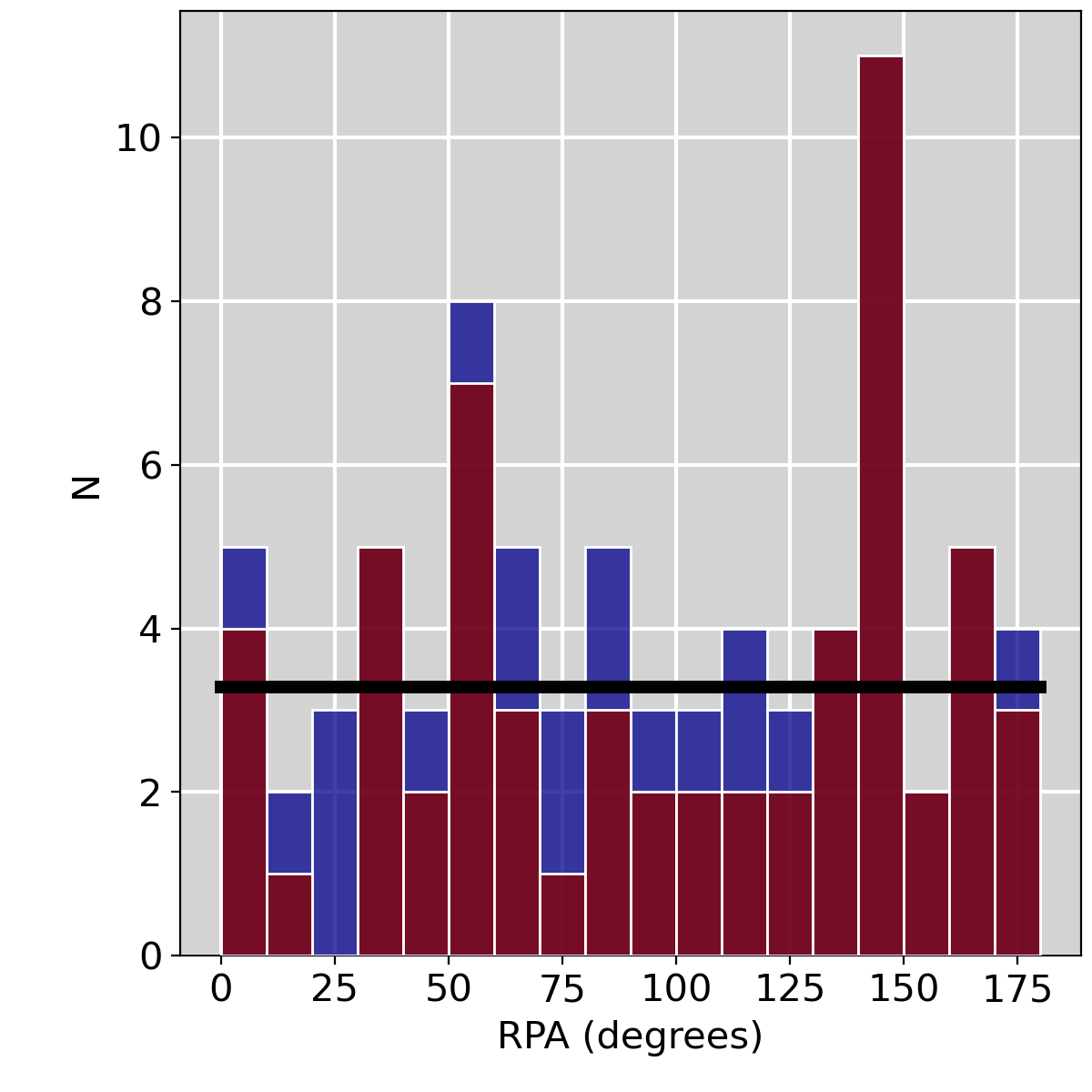}
        \caption{Distribution of the RPAs of the 78 ERGs (blue histogram) that we found in the inner region of the ELDF and of the 59 certain sources (red histogram). The black line shows the expected number of objects per bin for a uniform distribution of 78 ERGs.}
        \label{fig:Taylor_PA_distribution}
  \end{figure}

\begin{figure}[]
        \centering
        \includegraphics[width= 0.50\textwidth]{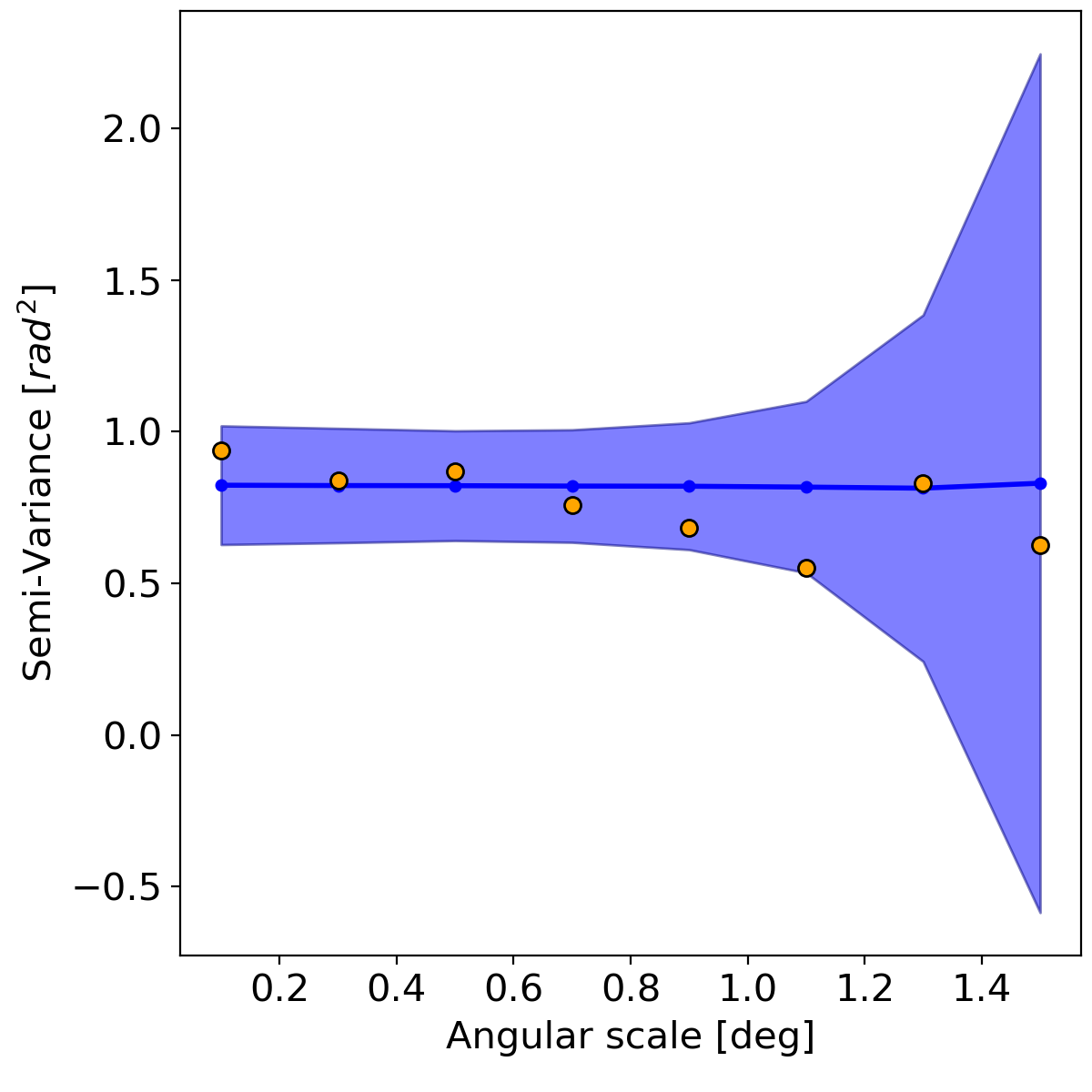}
        \caption{Estimate of the semi-variance on different angular scales in the inner region of the ELAIS-N1 field. The blue line and points are the semi-variance values obtained for randomly generated position angles with the same spatial distribution of the 78 ERGs. The shadowed region represents the $2\sigma_{\rm SM}$ values. The orange points are estimated from our dataset.}
        \label{fig:inner_region_semi_variance}
\end{figure}

Here, we look at the distribution of the RPAs in the inner $\sim$1.7 $\rm deg^2$ of the ELAIS-N1 field ($241.5 < RA < 243.75, 53.9 < DEC < 55.2$), where \citet{Taylor2016} found a statistically significant alignment of radio galaxies. We recall that 9 radio sources they used in their analysis are not actual radio galaxies and we could add 24 more ERGs. Thus, the sample for such analysis consists of 78 ERGs, of which 19 are flagged as uncertain RPA measurement. We show the distribution of the RPAs in the inner region of the ELAIS-N1 field in Fig.~\ref{fig:Taylor_PA_distribution}. The blue histogram shows the distribution of the total sample of RPA in this region, while in the red histogram the uncertain measurements are excluded. The figure clearly shows a peak at RPAs around 140$^\circ$ in agreement with \citet{Taylor2016}. We then carried out the statistical tests explained in Sec.~\ref{sec:tests} and found a p-value of 0.66 and 0.31 for the KS test and the $\chi^2$ test, respectively. The latter test is valid for large samples and it is customary to recommend, in applications of the test, that the smallest expected number in any bin should be 5 \citep{Cochran1952}. We performed the test using 13 bins with a width of 15$^\circ$ which lead to an expected value of about 6.5 elements per bin. The resulting p-value, in this case, is 0.23. 
Concerning the Rayleigh test, we found a mean resultant length $\bar{R} = 0.009$ which results in a p-value=0.96. Thus, even though the distribution shows a clear peak, we cannot reject the hypothesis of uniformity of the RPAs in this region. Moreover, the analysis involving the semi-variance (Fig.~\ref{fig:inner_region_semi_variance}) shows that there is no correlation between the RPAs of the ERGs, located at different positions of the sky, at any angular scale. Here, the blue line and points are the values estimated by using randomly generated data which have the same spatial distribution of the 78 ERGs in the inner region of the ELAIS-N1 field, while the orange points are the result of the analysis performed on our dataset. \newline
We did not perform an analysis based on the dispersion measure (that is the 5th method listed in Sec.~\ref{sec:tests}) due to the smaller number of ERGs when restricting the study to the inner region of the field. As a matter of fact, with a sample of only 78 objects, we are certainly dominated by the shot noise \citep{Jain2004} which would cancel out any signal unless the alignment is very strong, which does not seem to be the case here. \newline
We performed the statistical tests on the sample of 59 ERGs for which we could measure a reliable RPA as well. We obtained a p-value of 0.10, 0.01 and 0.46 for the KS, $\chi^2$ and Rayleigh test, respectively. The result of the $\chi^2$ test holds when considering bins with a width of 15$^\circ$. Nevertheless, this is the only test which suggests an alignment of the ERGs in the inner region as also the semi-variance test applied to this smaller sample cannot reject the hypothesis of a uniform distribution.\newline
The sensitivity of the LOFAR (20 $\mu Jy/$beam)  and GMRT (10 $\mu Jy/$beam) ELAIS-N1 deep field observations are quite similar, but the four times lower frequency of LOFAR makes a RG, with a typical spectral index of -0.8, about three times brighter at 144 MHz compared to 610 MHz. Moreover, the availability of deeper infrared source catalogues like CatWISE \citep{Marocco2021} and unWISE \citep{unWISE2019} enables the identification of more distant galaxies which may emit in the radio band as point-like sources. Such contamination, if superimposed on the emission of an ERG, may slightly change the morphology of the latter and lead to a wrong RPA measurement.\newline
In order to attempt to reproduce the \citet{Taylor2016} results, we extracted the positions, sizes and RPAs of the RGs from their Fig.~2 as follows:
the end points of all vectors were digitized with the g3data software,
and saved as RA, DEC in degrees. Then, we reviewed the RPA measurements and could closely match the histogram shown in their Fig.~3. We ran our first four statistical tests on the recovered data, but found that none of them is able to reject the hypothesis of uniformity. In particular, for the Rayleigh test, we obtained a mean resultant length of 0.09 from our analysis of these data, which is highly discrepant from the value of 0.68 derived by \citet{Taylor2016} that led them to conclude non-uniformity of RPAs. The origin of this difference is uncertain, although we note that if we omit to multiply the RPAs by a factor of two (a step which is required, since the test assesses uniformity over a circle and the RPAs are distributed over $\left[0,180 \right)$) then we obtain an erroneous mean resultant length of 0.64, which is much closer to the value quoted by \citet{Taylor2016}.

\subsection{Alignment in the entire ELAIS-N1 field}
\label{sec:total_alignment}

\begin{figure}[]
        \centering
        \includegraphics[width= 0.50\textwidth]{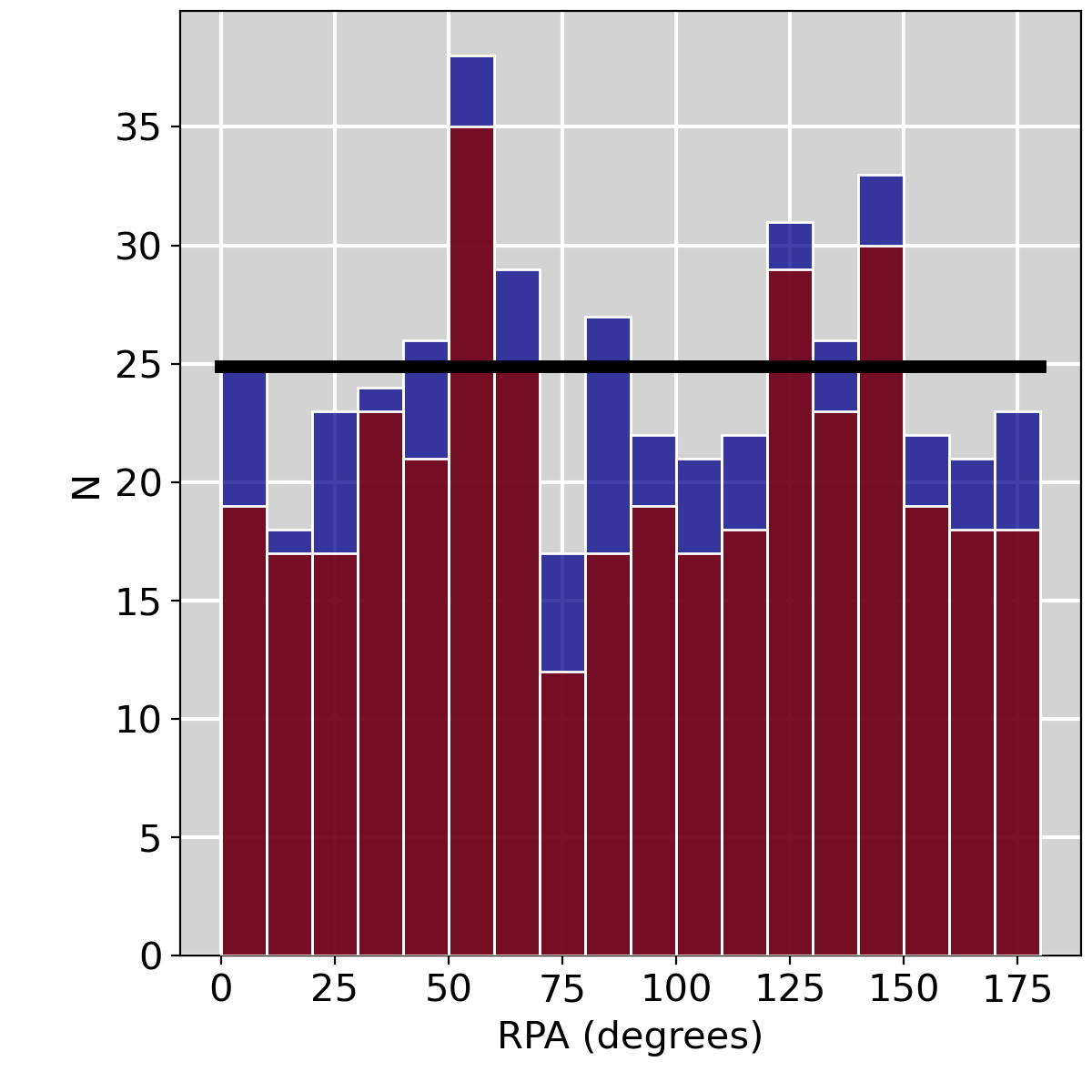}
        \caption{Distribution of the RPAs of the 447 ERGs we found in the ELDF (red histograms) and of the 377 certain sources (red histogram). The black line shows the expected number of objects per bin for a uniform distribution considering the total sample.}
        \label{fig:PA_distribution}
\end{figure}

\begin{figure}[]
        \centering
        \includegraphics[width= 0.50\textwidth]{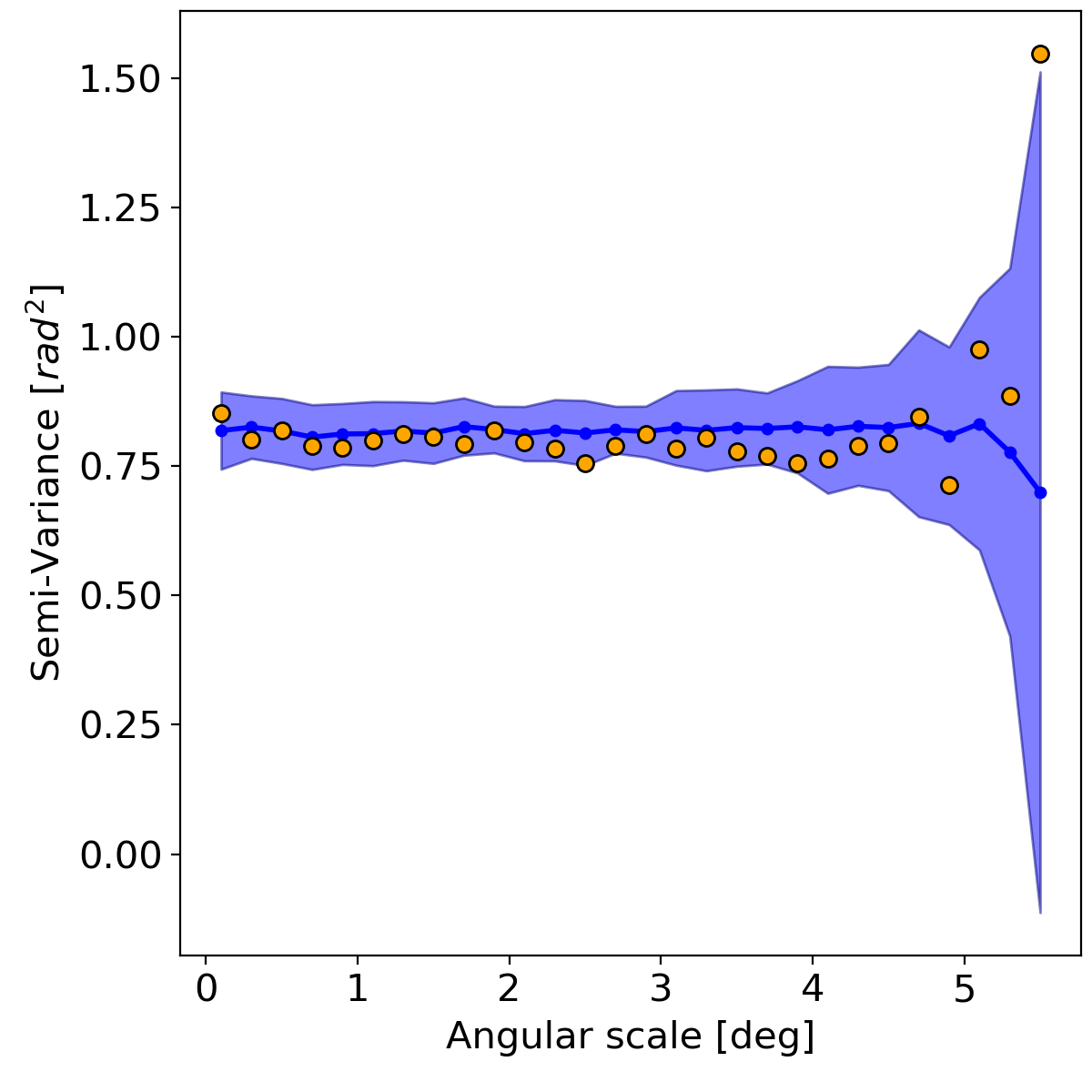}
        \caption{Estimate of the semi-variance on different angular scales. The blue line and points highlight the constant value of the semi-variance for randomly generated position angles with the same spatial distribution of the 447 ERGs in our sample. The shadowed region represents the $2\sigma_{\rm SM}$ values. The orange points are the semi-variance values of our sample.}
        \label{fig:semi_variance}
\end{figure}

We show the distribution of the RPAs of the radio galaxies in the ELDF in Fig.~\ref{fig:PA_distribution}. The blue histogram represents the total sample, while the red histogram shows the distribution for the 377 certain sources, i.e. those ERGs that do not show a complex morphology and for which we could accurately measure the RPA. The black line denotes the expected number of objects per bin if the distribution were uniform. Now, we performed the same statistical tests considering the total sample. The results, with a p-value equal to 0.71, 0.33 and 0.88 for the KS test, $\chi^2$ test and Rayleigh test respectively, suggest that the uniformity holds when including the entire field as well. 
These results are also confirmed by the analysis of the semi-variance. We measured the semi-variance in our sample, shown by the orange points in Fig.~\ref{fig:semi_variance}. The blue line and points are the semi-variance values estimated from randomly generated data and the shadowed region represents the $2\sigma_{\rm SM}$ values. The larger uncertainties on the largest scale are due to poor statistics since not many pairs are separated by such large distances. Overall, there is no clear evidence for a convincing signal in favour of an alignment as the orange points are always consistent with 0.82 within the error. \newline

\begin{figure}[]
        \centering
        \includegraphics[width= 0.50\textwidth]{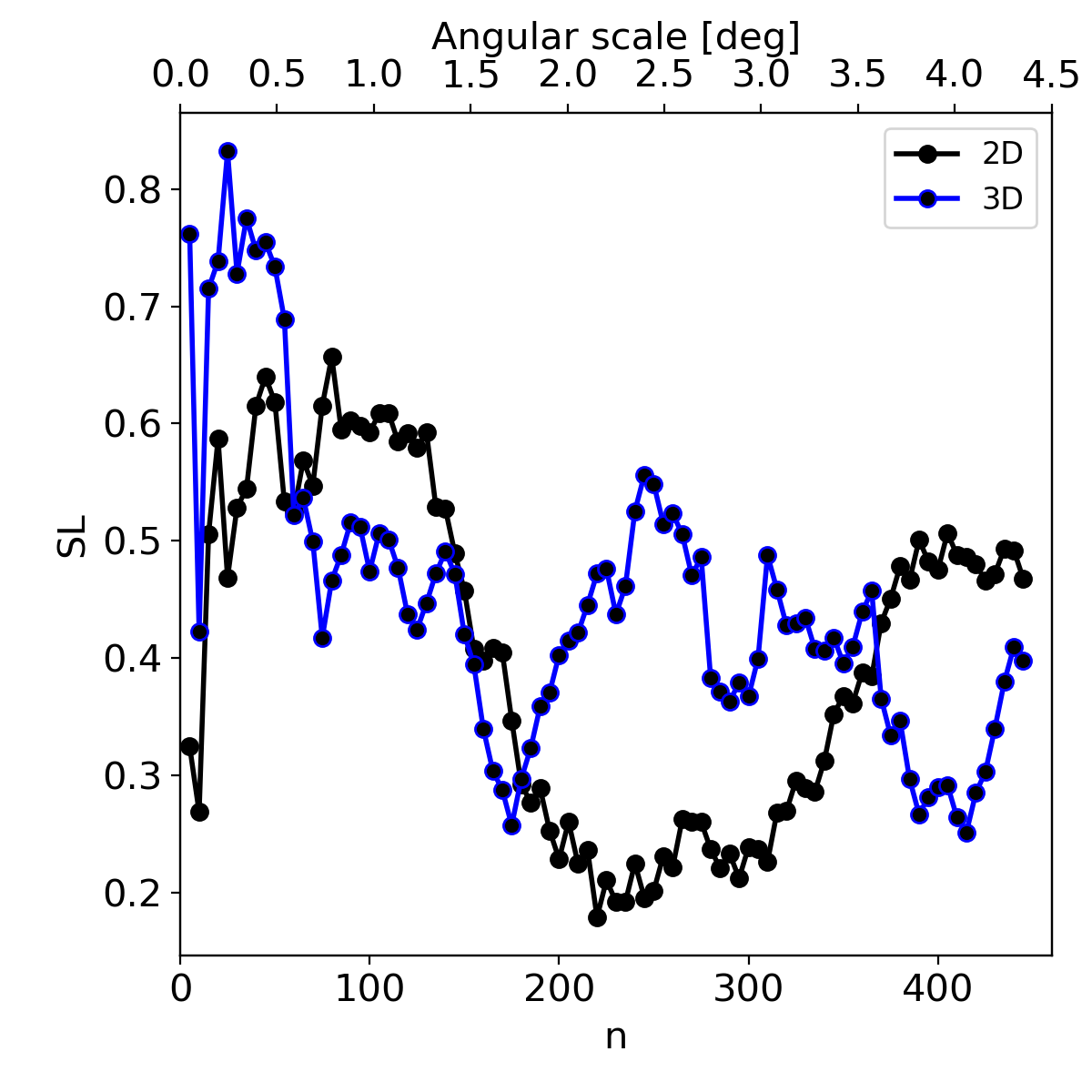}
        \caption{Significance level of the dispersion measure test (SL) as a function of the nearest neighbours (n, lower abscissa) and angular scale in degrees (upper abscissa). The black line shows the results of the 2D analysis while the 3D analysis is shown with the blue line. Such a large SL ($>>0.03$) shows that no alignment is present in the ELAIS-N1 field at any scale in our analysis.}
        \label{fig:significance_level}
\end{figure}

\begin{figure}[]
        \centering
        \includegraphics[width= 0.50\textwidth]{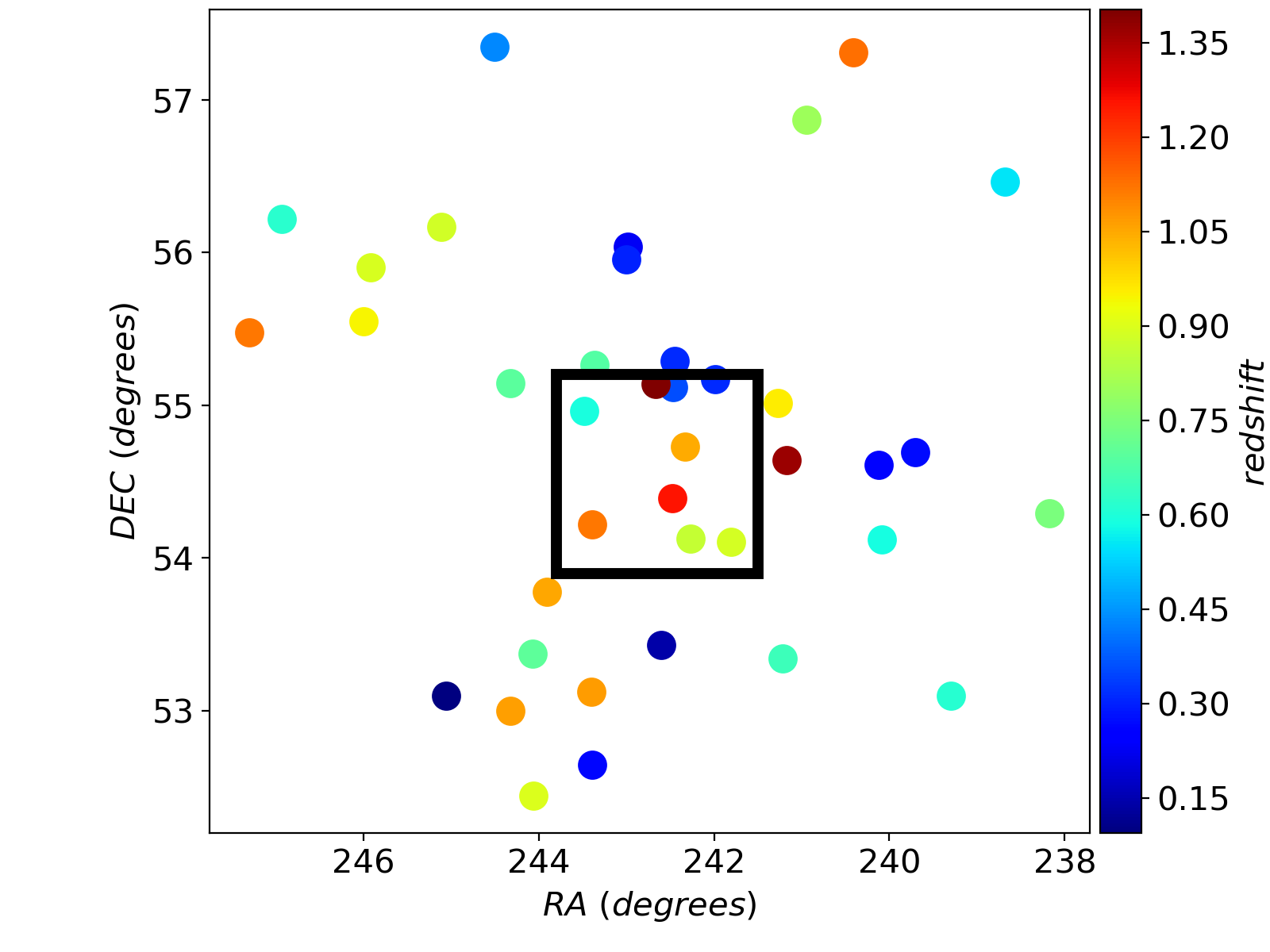}\quad\includegraphics[width= 0.50\textwidth]{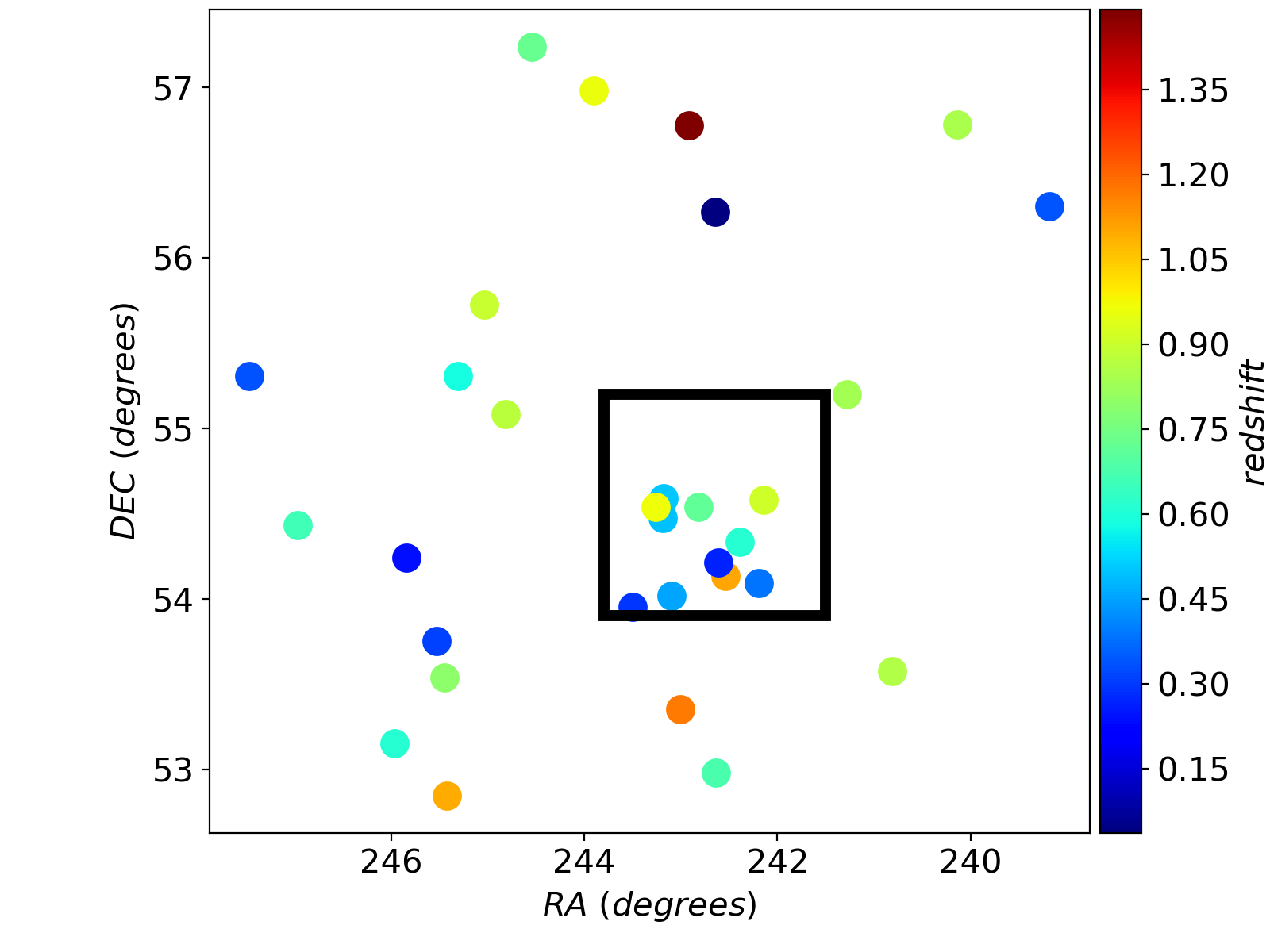}
        \caption{Spatial distribution of 39 ERGs with RPA between $50^{\circ}$ and $60^{\circ}$ (upper panel) and 31 ERGs with RPA in the range $140^{\circ}-150^{\circ}$. The colorbar shows the redshifts of the ERGs. The analysis was restricted to $z \le 1.5$. The black box represents the field of \cite{Taylor2016}.}
        \label{fig:RGs_distribution}
\end{figure}

Finally, we show the results of the 2D (black line) and 3D (blue line) dispersion measure tests in Fig.~\ref{fig:significance_level}. The significance level, $SL$, is plotted as a function of the number of nearest neighbours, n, and angular scale in degrees. Following previous studies \citep[e.g.,][]{Contigiani2017}, a commonly used criterion for the presence of an alignment signal is $SL<$0.03 ($\rm log(SL)<-1.5)$. As mentioned in Sec.~\ref{sec:tests}, we are more affected by the shot noise due to the comparatively smaller size of our sample. However, a minimum significance level of about 0.2 in Fig.~\ref{fig:significance_level} suggests there is no evident signal, neither in the 2D nor in the 3D analysis, at any scale. \newline
These results also hold when considering only the ERGs with reliable RPA measurement.\newline
Even though the tests suggest that radio galaxies are randomly oriented, two conspicuous peaks are visible on an RPA range between $50^{\circ}-60^{\circ}$ and $\sim 140^{\circ}-150^{\circ}$ (the latter was seen by \citealt{Taylor2016} as well). The Poisson distribution gives the probability that a given number of observations fall within an interval of values knowing the average frequency of that particular event. Thus, using such a distribution we find that the two peaks are $\sim2.5\sigma$ (for RPAs between $ 50^{\circ}-60^{\circ}$) and $\sim1.5\sigma$ (for RPA between $140^{\circ}-150^{\circ}$) above the average. In Fig.~\ref{fig:RGs_distribution} we show the spatial and redshift distributions of the ERGs with an orientation between 50$^{\circ}$-60$^{\circ}$ (upper panel) and 140$^{\circ}$-150$^{\circ}$ (lower panel). We selected the ERGs up to redshift 1.5 since the majority of ERGs at larger redshifts either do not have a redshift estimate in the literature or have very large errors. The black rectangles highlight the region inspected by \citet{Taylor2016}. In both cases, there is no 3D alignment of ERGs as the redshifts span a range from  $0.1 \lesssim z \lesssim 1.5$. 

\section{Discussion and summary}
\label{sec:discussion}

The tidal torque theory predicts that the angular momentum of the dark matter proto-halos is acquired during their formation which occurs along the entire evolution of the large-scale structure of the universe \citep{Peebles1969, Doroshkevich1970, White1984, Porciani2002, Schafer2009}. As a result, an alignment between optical galaxies and the large-scale structure (e.g. filaments and sheets) is expected \citep{Hu2006,Joachimi2015,Kirk2015}. In a first attempt to study this alignment \citet{Hawley1975} found a small departure from isotropy in the distribution of the orientation angle, measured as the angle between the major axis of the galaxy and the local meridian. \citet{Lee2004} argued that the observed large-scale coherence in the orientation of nearby spiral galaxies found by \citet{Navarro2004} can be fully explained by the tidal torque theory. Others have tried to look at a possible alignment of galaxies and most of these found that the minor axes of early-type galaxies are preferentially oriented perpendicular to the host filament \citep{Tempel2013a, Tempel2013b,Hirv2017}, while late-type galaxies have spin axes parallel to the closest filament \citep{Tempel2013a,Tempel2013b,Hirv2017,Bird2020,Kraljic2021,Tudorache2022}. However, some conflicting results have been found \citep{Jones2010,Zhang2015,Pahwa2016,Krolewski2019}. Recently, \citet{Rodriguez2022}, by using the IllustrisTNG simulations \citep{Nelson2019}, found an alignment with the large-scale structure of red galaxies in the centres of galaxy clusters and groups. They then speculated that this anisotropy in the orientation of the central galaxies is the consequence of a concatenation of alignments. Starting from the alignment between the central galaxy and the host cluster \citep{Yuan2022}, eventually, the host halo aligns with the structures surrounding it. \newline
Some work found that there is a mild preference for radio jets to align with the minor axis of the galaxy host \citep{Kotanyi1979,Battye2009,Kamali2009, Vazquez2019}. Assuming that the alignment between radio jets and optical galaxies is real, one could in principle look at the alignment between the radio galaxies and the large-scale structure \citep[e.g.,][]{West1991}. Nevertheless, some opposing results regarding the orientations of radio jets have been found \citep{Schmitt2002,Klejin2005,Hopkins2012} casting doubts on this assumption. \newline
In this work, we revisited the alignment of radio jets in the ELAIS-N1 field. We inspected the LOFAR ELAIS-N1 deep field in which we identified the host galaxies of 447 ERGs whose radio emission extends over at least $\sim 0.5'$. We measured the RPA of the major radio axis (assuming it is a tracer of the underlying radio jets direction) and studied their distribution by using a number of statistical tests, none of which is able to reject the null hypothesis of uniform orientations. Similar results are obtained when restricting the analysis to the region inspected by \citet{Taylor2016}. Only when restricting the sample to the 59 ERGs with reliable RPA measurement in the inner region, the $\chi^2$ test returns a p-value=0.01 (i.e. it attributes a 1\% chance of the result being a statistical fluctuation). However, none of the other statistical tests on this sample is able to reject the hypothesis of uniformity of the RPA distribution. We recovered the data used by \citet{Taylor2016} for their analysis and showed that, even with such sample, we could not obtain the same results. Furthermore, we found that the redshifts of ERGs with orientations near the two peaks (around 50$^{\circ}$ and 140$^{\circ}$) span a wide range, $0.1 \lesssim z \lesssim 1.5$, strongly arguing
against the idea of a 3D alignment of radio galaxies. Other reports of a 3D alignment \citep[e.g.,][]{Contigiani2017, Panwar2020} have not been statistically significant. However, several studies reported a 2D alignment \citep{Contigiani2017, Panwar2020, Mandarakas2021} over angular scales similar to those that we studied. The maximum angular scale we could explore is $\sim 4^\circ$ (see Fig.~\ref{fig:semi_variance}) which is the scale over which \citet{Osinga2020} found a 2D alignment. The combination of the two results might suggest that the 2D alignment of radio galaxies may exist on scales larger than those probed by our analysis. \newline
  
\section*{Acknowledgements}
{\small This work is funded by the Deutsche Forschungsgemeinschaft (DFG, German Research Foundation) under Germany's Excellence Strategy -- EXC 2121 ``Quantum Universe'' --  390833306 as well as grant DFG BR2026/27.

HA has benefited from grant CIIC\,138/2022 of Universidad de Guanajuato, Mexico. 

PNB is grateful for support from the UK STFC via grant ST/V000594/1.

EO acknowledges support from the VIDI research programme with project number 639.042.729

LOFAR \citep{Lofar2013} is the Low Frequency Array designed and constructed by ASTRON. It has observing, data processing, and data storage facilities in several countries, which are owned by various parties (each with their own funding sources), and that are collectively operated by the ILT foundation under a joint scientific policy. The ILT resources have benefited from the following recent major funding sources: CNRS-INSU, Observatoire de Paris and Université d'Orléans, France; BMBF, MIWF-NRW, MPG, Germany; Science Foundation Ireland (SFI), Department of Business, Enterprise and Innovation (DBEI), Ireland; NWO, The Netherlands; The Science and Technology Facilities Council, UK; Ministry of Science and Higher Education, Poland; The Istituto Nazionale di Astrofisica (INAF), Italy.

This research made use of the Dutch national e-infrastructure with support of the SURF Cooperative (e-infra 180169) and the LOFAR e-infra group. The Jülich LOFAR Long Term Archive and the German LOFAR network are both coordinated and operated by the Jülich Supercomputing Centre (JSC), and computing resources on the supercomputer JUWELS at JSC were provided by the Gauss Centre for Supercomputing e.V. (grant CHTB00) through the John von Neumann Institute for Computing (NIC).
This research made use of the University of Hertfordshire high-performance computing facility and the LOFAR-UK computing facility located at the University of Hertfordshire and supported by STFC [ST/P000096/1], and of the Italian LOFAR IT computing infrastructure supported and operated by INAF, and by the Physics Department of Turin university (under an agreement with Consorzio Interuniversitario per la Fisica Spaziale) at the C3S Supercomputing Centre, Italy.
}



\bibliography{ref}
\bibliographystyle{aa}

\begin{appendix}

\section{Notes on discarded sources}
\label{appendix}

J1609+5452 ($RA=242.4899,DEC=54.8794$). The radio emission coming from the two hotspots of this FRII candidate is the result of  two separate point-like radio sources, hosted by SDSS J160958.85+545249.4 and CWISE J160956.26+545243.4.\newline

J1609+5500 ($RA=242.4858,DEC=55.0000$) is actually two separate sources. The source due to NE is likely a RG itself (which we did not include in our sample) with uncertain host galaxy. We propose CWISE J160959.21+550019.8 as the host candidate, even though the radio morphology makes this claim uncertain. On the other hand, The SW source is a clear Wide Angle Tailed (WAT) RG with host galaxy SDSS J160952.46+545937.9.\newline

J1610+5416 ($RA=242.5074,DEC=54.2767$) might be a genuine RG with host SDSS J161002.72+541640.0. However, the radio morphology is very unclear and we consider the RPA measurement of this source very uncertain. Thus, following the criteria applied during our inspection, we discarded this source.  \newline

J1610+5506 ($RA=242.5574,DEC=55.1125$) is a candidate RG whose radio emission is the result of at least two separate sources, hosted by SDSS J161013.41+550649.9 and SDSS J161014.30+550638.3.\newline

J1612+5358 ($RA=243.1542,DEC=53.9686$) is the spiral galaxy SDSS J161236.98+535807.2. Thus, the RPA of the radio source is just the major axis of the galaxy and not that of any jets.\newline

J1613+5411 ($RA=243.3983,DEC=54.1908$) is likely the combination of two different radio emissions. Part of the emission comes from the radio lobe of J1613+5412 an ERG (with host CWISE J161336.25+541258.4) that is included in our sample and whose RPA measurement is reliable. Moreover, there might be a contribution coming from the host galaxy SDSS J161334.98+541112.0 for which a reliable RPA measurement is not possible. \newline

J1613+5414a ($RA=243.4417,DEC=54.2347$) is the NE lobe of another ERG (host galaxy CWISE J161334.24+541304.3) we have in our sample used for the analysis. Again, there might be a contribution from another point-like source with host galaxy SDSS J161347.93+541405.6. \newline

J1613+5414b ($RA=243.3708,DEC=54.2489$) it is likely the result of the emission of two separate sources: DESI J243.3650+54.2490 and CWISE J161329.52+541454.7.\newline

J1613+5415 ($RA=243.4104,DEC=54.2514$) is the N lobe of J1613+5412.

\begin{table*}
\centering
\caption{The first 50 rows of the ERGs catalogue we used for our analysis.}
    \begin{threeparttable}
    \begin{tabular}{p{1.6cm}p{1.2cm}p{1cm}p{0.3cm}R{0.7cm}p{0.5cm}p{0.5cm}p{0.4cm}p{0.5cm}p{4.4cm}p{0.5cm}p{1.1cm}p{0.7cm}}
            \hline
                     (1) & (2) & (3) & (4) & (5) & (6) & (7) & (8) & (9) & (10) & (11) & (12) & (13) \\
                    Name & RA$_J$ & ${\rm Dec}_J$ & \textit{LAS}  & RPA & FR & $z$  & $\Delta z$ & ztype & Hostname & type & mag & \textit{LLS} \\ 
                         & $^{\circ}$ & $^{\circ}$ & ($'$) &$^{\circ}$ & type & & &  & & & (mag) & (Mpc) \\ \hline
                
                J1552+5540 & 238.1360 & 55.6731 & 0.50 & 10 & II & 1.50 & ~ & e & CWISE J155232.64+554023.1 & Qc & 18.60W1 & 0.25 \\ 
        J1552+5417 & 238.1721 & 54.2920 & 0.68 & 50 & II & 0.74 & 0.06 & p & SDSS J155241.30+541731.3 & G & 21.25r & 0.30 \\ 
        J1553+5454 & 238.2511 & 54.8994 & 1.23 & 30 & II & 0.97 & 0.23 & p & DESI J238.2511+54.8994 & Qc & 23.67r & 0.59 \\ 
        J1553+5516 & 238.3149 & 55.2773 & 0.45 & 155 & II & 0.88 & 0.04 & p & DESI J238.3149+55.2773 & G & 23.11r & 0.21 \\ 
        J1554+5506 & 238.5454 & 55.1050 & 1.31 & 67 & II & 0.720 & ~ & s & SDSS J155410.89+550617.9 & G & 22.11r & 0.57 \\ 
        J1554+5438 & 238.5471 & 54.6374 & 0.93 & 44 & I/II & 0.925 & ~ & s & SDSS J155411.32+543814.8 & G & 21.09r & 0.44 \\ 
        J1554+5334 & 238.6527 & 53.5750 & 2.55 & 25 & II & 0.761 & ~ & s & SDSS J155436.64+533430.0 & G & 21.61r & 1.13 \\ 
        J1554+5344 & 238.6665 & 53.7407 & 1.08 & 84 & II & 0.74 & 0.04 & p & SDSS J155439.96+534426.5 & G & 22.07r & 0.47 \\ 
        J1554+5628 & 238.6835 & 56.4664 & 0.72 & 55 & II & 0.554 & ~ & s & SDSS J155444.03+562758.9 & G & 20.85r & 0.28 \\ 
        J1556+5539 & 239.0106 & 55.6599 & 2.25 & 87 & II & 0.31 & 0.04 & p & SDSS J155602.54+553935.6 & G & 18.80r' & 0.62 \\ 
        J1556+5348 & 239.0622 & 53.8046 & 0.58 & 10 & II & 0.98 & 0.11 & p & DESI J239.0622+53.8046 & G & 23.12r & 0.28 \\ 
        J1556+5559 & 239.0914 & 55.9918 & 0.55 & 73 & II & 0.97 & 0.14 & p & SDSS J155621.93+555930.5 & G & 22.55r & 0.26 \\ 
        J1556+5627 & 239.1139 & 56.4524 & 0.63 & 33 & II & 0.75 & 0.02 & p & SDSS J155627.34+562708.8 & G & 21.04r & 0.28 \\ 
        J1556+5538 & 239.1143 & 55.6493 & 0.48 & 179 & II & 0.91 & 0.08 & p & SDSS J155627.42+553857.4 & G & 22.67r & 0.22 \\ 
        J1556+5342 & 239.1211 & 53.7064 & 0.60 & 24 & II & 0.71 & 0.03 & p & SDSS J155629.05+534222.9 & G & 22.63r' & 0.26 \\ 
        J1556+5430 & 239.1267 & 54.5048 & 0.72 & 177 & II & 1.12 & 0.01 & p & DESI J239.1267+54.5048 & G & 23.28r & 0.35 \\ 
        J1556+5353 & 239.1312 & 53.9000 & 0.78 & 77 & II & 0.87 & 0.12 & p & DESI J239.1312+53.9000 & G & 23.31r & 0.36 \\ 
        J1556+5314 & 239.1368 & 53.2427 & 2.63 & 156 & II & 0.241 & ~ & s & SDSS J155632.82+531433.7 & G & 17.71r & 0.60 \\ 
        J1556+5618 & 239.1805 & 56.2996 & 1.38 & 148 & II & 0.340 & ~ & s & SDSS J155643.33+561758.7 & G & 18.21r & 0.40 \\ 
        J1556+5407 & 239.1856 & 54.1247 & 1.05 & 152 & II & 0.499 & ~ & s & SDSS J155644.54+540728.8 & G & 20.10 & 0.38 \\ 
        J1556+5558 & 239.1902 & 55.9771 & 0.50 & 178 & II & 0.83 & 0.22 & p & DESI J239.1902+55.9771 & G & 23.18r & 0.23 \\ 
        J1556+5422 & 239.2119 & 54.3713 & 1.12 & 40 & II & 1.50 & ~ & e & CWISE J155650.86+542216.8 & Qc & 17.01W1 & 0.57 \\ 
        J1556+5647 & 239.2142 & 56.7877 & 0.67 & 30 & II & 0.91 & 0.08 & p & DESI J239.2142+56.7877 & G & 22.58r & 0.31 \\ 
        J1557+5446 & 239.2544 & 54.7722 & 0.70 & 122 & I & 0.73 & 0.07 & p & DESI J239.2544+54.7722 & G & 22.27r & 0.30 \\ 
        J1557+5305 & 239.3013 & 53.0949 & 1.24 & 50 & II & 0.61 & 0.03 & p & SDSS J155712.32+530541.5 & G & 20.85r & 0.50 \\ 
        J1557+5442 & 239.3131 & 54.7131 & 0.58 & 90 & II & 0.928 & ~ & s & SDSS J155715.14+544247.2 & G & 22.31r' & 0.27 \\ 
        J1557+5409 & 239.3297 & 54.1571 & 1.87 & 122 & II & 0.81 & 0.32 & p & DESI J239.3297+54.1571 & G & 22.85r & 0.85 \\ 
        J1557+5440 & 239.3392 & 54.6711 & 11.20 & 154 & I & 0.047 & ~ & s & SDSS J155721.39+544015.9 & G & 13.95r' & 0.62 \\ 
        J1557+5448 & 239.3770 & 54.8036 & 1.08 & 34 & II & 0.76 & 0.28 & p & PSO J155730.480+544812.94 & G & 20.68r & 0.48 \\ 
        J1557+5402 & 239.4011 & 54.0334 & 0.50 & 11 & II & 1.06 & 0.06 & p & DESI J239.4011+54.0334 & G & 24.37r & 0.24 \\ 
        J1557+5622 & 239.4081 & 56.3710 & 0.70 & 178 & II & 1.50 & ~ & e & CWISE J155737.93+562215.6 & Qc & 17.94W1 & 0.36 \\ 
        J1557+5527 & 239.4508 & 55.4624 & 0.83 & 122 & I & 0.462 & ~ & s & SDSS J155748.18+552744.5 & G & 19.69r' & 0.29 \\ 
        J1557+5327 & 239.4540 & 53.4701 & 1.64 & 73: & I/II & 0.67 & 0.02 & p & SDSS J155748.96+532812.3 & G & 21.39r' & 0.69 \\ 
        J1557+5508 & 239.4560 & 55.1401 & 0.65 & 111 & I/II & 1.90 & ~ & e & CWISE J155749.44+550824.5 & Qc & 19.04W1 & 0.33 \\ 
        J1557+5343 & 239.4594 & 53.7261 & 0.83 & 100 & II & 0.312 & ~ & s & SDSS J155750.24+534334.0 & G & 17.82r & 0.23 \\ 
        J1558+5403 & 239.5270 & 54.0549 & 0.33 & 137 & II & 1.03 & 0.12 & p & DESI J239.5270+54.0549 & G & 23.51r & 0.16 \\ 
        J1558+5554 & 239.5512 & 55.9047 & 0.68 & 68 & II & 1.21 & 0.19 & p & DESI J239.5512+55.9047 & G & 25.00r & 0.34 \\ 
        J1558+5633 & 239.5623 & 56.5503 & 0.63 & 92 & II & 0.92 & 0.01 & p & DESI J239.5623+56.5503 & G & 23.42r & 0.30 \\ 
        J1558+5558 & 239.5700 & 55.9836 & 0.85 & 48 & II & 0.50 & 0.09 & p & SDSS J155816.81+555900.9 & G & 21.91r & 0.31 \\ 
        J1558+5349 & 239.5759 & 53.8298 & 0.50 & 33 & II & 1.03 & 0.16 & p & DESI J239.5759+53.8298 & Qc & 23.46r & 0.24 \\ 
        J1558+5303 & 239.6051 & 53.0636 & 1.77 & 40 & I & 0.687 & ~ & s & SDSS J155825.22+530348.8 & G & 20.94r & 0.75 \\ 
        J1558+5609 & 239.7018 & 56.1579 & 0.45 & 115 & I/I & 1.28 & ~ & p & DESI J239.7018+56.1579 & Qc & 19.84r & 0.23 \\ 
        J1558+5441 & 239.7042 & 54.6902 & 1.08 & 52 & I & 0.27 & 0.03 & p & SDSS J155849.00+544124.8 & G & 18.36r & 0.27 \\ 
        J1558+5317 & 239.7399 & 53.2952 & 0.53 & 161 & II & 0.472 & ~ & s & SDSS J155857.57+531742.6 & G & 20.13r & 0.19 \\ 
        J1559+5527 & 239.7540 & 55.4503 & 0.40 & 136 & II & 1.16 & 0.64 & p & SDSS J155900.96+552701.0 & G & 22.86r' & 0.20 \\ 
        J1559+5704 & 239.7827 & 57.0779 & 1.30 & 172 & II & 0.592 & ~ & s & SDSS J155907.84+570440.4 & G & 20.31r & 0.52 \\ 
        J1559+5448 & 239.8707 & 54.8169 & 0.70 & 5: & II & 0.85 & 1.17 & p & SDSS J155928.96+544900.8 & G & 19.86r & 0.32 \\ 
        J1559+5341 & 239.9018 & 53.6960 & 0.90 & 165 & II & 1.09 & 1.06 & p & DESI J239.9018+53.6960 & G & 21.32r & 0.44 \\ 
        J1559+5708 & 239.9340 & 57.1458 & 0.87 & 113 & II & 0.50 & 0.01 & p & SDSS J155944.17+570844.9 & G & 20.84r & 0.32 \\ 
        J1559+5556 & 239.9753 & 55.9342 & 0.38 & 31 & II & 1.50 & ~ & e & CWISE J155954.08+555603.2 & Qc & 18.03W1 & 0.19 \\ \hline

        \label{tab:ERGs_list}
\end{tabular}
\begin{tablenotes}
      \item[(a)] \small{Col.~(1), name of the ERG. Col.~(2) and Col.~(3), right ascension and declination (J2000) of the host galaxy in degrees. Col.~(4), largest angular size in arcminute. Col.~(5) Radio position angle of the ERG in degrees; a colon appended to the RPA indicates that the measurement is uncertain. Col.~(6), classification of the ERG according to the Fanaroff–Riley classification. Col.~(7), redshift of the host galaxy. Col.~(8), redshift error when available. Errors of the spectroscopic redshifts are not reported since they are generally more accurate than the precision we can achieve on the linear sizes. Col.~(9), type of the redshift: p for photometric, s for spectroscopic and e if estimated. Col.~(10), name of the host galaxy. Col.~(11) type of host galaxy: galaxy (G) or QSO (Q) or candidate quasars (Qc). Col.~(12), magnitude of the host galaxy in the r-band if available from the DESI DR9 photometric catalogue (r) or SDSSDR12 (r'); the label W1 and W2 indicates that the magnitude is taken from either WISEA \citep{Cutri2012,Cutri2013} or CWISE \citep{Marocco2021} catalogue. Col.~(13), largest linear size in Mpc.}
\end{tablenotes}
\end{threeparttable}
\end{table*}

\end{appendix}

\end{document}